\documentclass{ifacconf}

\usepackage{graphicx}      
\usepackage{natbib}        
\usepackage{amsmath,amssymb,amsfonts,mathtools,bm}
\usepackage{booktabs}
\usepackage{subcaption}
\usepackage{algorithm}

\usepackage{algorithmic}

\newtheorem{assumption}{Assumption}
\newtheorem{remark}{Remark}

\newcommand{\tr}{\operatorname{Tr}}
\newcommand{\vecop}{\operatorname{vec}}
\newcommand{\argmin}{\operatorname*{arg\,min}}
\newcommand{\R}{\mathbb{R}}

\begin{document}
\begin{frontmatter}

\title{Influence Functions for Data Attribution in System Identification and LQR Control\thanksref{footnoteinfo}}

\thanks[footnoteinfo]{All authors are with the Department of Mechanical Engineering, The University of Texas at Austin, Austin, TX 78712, USA.}

\author[UT]{Jiachen Li}
\author[UT]{Shihao Li}
\author[UT]{Soovadeep Bakshi}
\author[UT]{Jiamin Xu}
\author[UT]{Dongmei Chen}

\address[UT]{Department of Mechanical Engineering, The University of Texas at Austin, Austin, TX 78712, USA
(e-mail: \{jiachenli, shihaoli01301, soovadeepbakshi, jiaminxu, dmchen\}@utexas.edu)}

\begin{abstract}
When a controller is designed from an identified model, its performance ultimately depends on the trajectories used for identification—but pinpointing which ones help or hurt remains an open problem. We bring influence functions, a data-attribution tool from machine learning, into this setting by chaining two closed-form sensitivity analyzes across a regularized least-squares identification and an infinite-horizon LQR pipeline. On the identification side, the quadratic loss admits an exact leave-one-trajectory-out (LOTO) parameter shift; a reusable first-order approximation follows with a Neumann-series error bound. On the control side, we implicitly differentiate through the DARE via its discrete Lyapunov structure and compress the cost gradient to a single adjoint Lyapunov solve. The resulting scores track true LOTO retraining with Pearson correlations above $0.99$ and $7$--$60\times$ speedups on linear systems of dimension $2$--$10$.
\end{abstract}


\end{frontmatter}

\section{Introduction}

Model-based control identifies a dynamics model from data and synthesizes a controller from that model. In safety-critical domains—autonomous vehicles, robotic manipulation, process control—the controller's performance traces back to the training data. A natural question follows: which trajectories matter most, and would the controller improve if certain ones were dropped?

Influence functions, rooted in deletion diagnostics \cite{cook1982residuals,cook1977detection} and extended to modern machine learning \cite{koh2017understanding}, answer such questions without retraining by inverting the training Hessian to approximate how a single example shifts learned parameters or a downstream loss. The technique has been applied to data debugging \cite{han2020explaining}, fairness auditing \cite{ghosh2023biased}, data valuation \cite{jia2019towards,wang2024shapley}, and large-model scaling \cite{li2024influence}—but always within prediction pipelines. Meanwhile, system identification \cite{ljung1999system,ljung2010perspectives}, experiment design for control \cite{hjalmarsson2005experiment}, and deep-learning-based identification \cite{pillonetto2025deep} address data collection and processing, not attribution. Riccati sensitivity has been studied for continuous-time \cite{konstantinov1993perturbation} and structured perturbations \cite{sun1998perturbation}, but not connected to data-level influence. DeePC \cite{coulson2019deepc} and related methods \cite{naf2025choose} bypass identification; differentiable MPC \cite{amos2018differentiable} enables gradient-based learning; SHAP has been applied to interpret MPC \cite{henkel2024interpretable}—yet none propagate trajectory-level attribution through an identification-to-controller pipeline. The core difficulty is that the map from model parameters to controller cost passes through the discrete algebraic Riccati equation (DARE).

\begin{figure}[t]
    \centering
    \includegraphics[width=8.4cm]{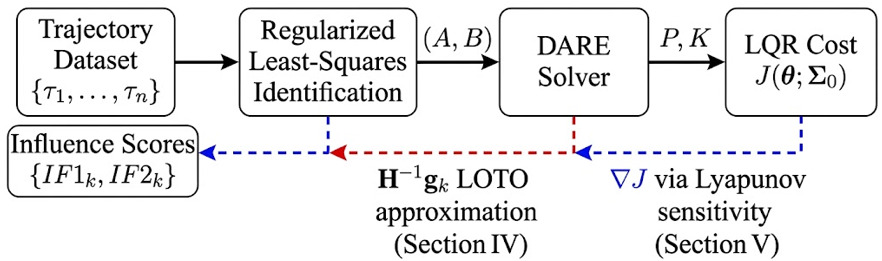}
    \caption{Overview of the compositional influence-function framework.}
    \label{fig:overview}
\end{figure}

We address this for the pipeline in Fig.~\ref{fig:overview}: Tikhonov-regularized least-squares identification followed by infinite-horizon LQR, chaining two analytically tractable sensitivity analyses. \textbf{Identification stage.} The regularized least-squares objective is quadratic, so the LOTO parameter shift admits an exact closed form; a first-order approximation with a Neumann-series error bound follows by dropping the per-trajectory curvature correction. \textbf{Control-design stage.} The Fr\'{e}chet derivative of the DARE residual reduces to a discrete Lyapunov operator, and an adjoint reformulation compresses the gradient computation from $p$ Lyapunov solves to one. Composing these stages yields end-to-end influence scores that match true LOTO retraining with Pearson correlations above $0.99$ and $7$--$60\times$ speedups on linear benchmarks of dimension $2$--$10$.

The linear setting makes both stages analytically tractable, isolating each source of approximation error. We also evaluate on a nonlinear two-link manipulator using a local linear surrogate; there the scores remain moderately informative (Pearson $0.620$), with degradation driven by the surrogate-to-plant gap rather than the influence approximation itself.

The paper is organized as follows. Section~2 formulates the pipeline. Section~3 develops the LOTO analysis for identification. Section~4 derives influence propagation through the DARE. Section~5 characterizes approximation errors. Section~6 presents empirical results. Section~7 concludes.

\section{Problem Formulation}

We consider a dataset of trajectories collected from an unknown discrete-time plant
\begin{equation}
    x_{t+1} = f_{\star}(x_t,u_t) + w_t,
    \label{eq:true_plant}
\end{equation}
where $x_t \in \R^{n_x}$, $u_t \in \R^{n_u}$, and $w_t$ are process disturbances. The plant may be nonlinear; all subsequent analyzes are carried out on a learned linear surrogate.

The training set is $\mathcal{D} = \{\tau_1,\dots,\tau_N\}$, where each trajectory $\tau_k = \{(x_s,u_s,x_s^+)\}_{s\in\mathcal{I}_k}$ consists of $T_k := |\mathcal{I}_k|$ transition tuples, with $x_s^+$ denoting the successor state.

\textbf{Identification stage.} We fit a linear surrogate
\begin{equation}
    \hat{x}_{t+1} = A_{\theta}x_t + B_{\theta}u_t,
    \label{eq:lin_surrogate}
\end{equation}
parameterized by $\theta = \vecop(A_{\theta},B_{\theta}) \in \R^p$ with $p = n_x(n_x + n_u)$. Adopting the regression form $\hat{x}_{t+1} = \Phi_s\theta$ where $\Phi_s := \Phi(x_s,u_s)\in\R^{n_x\times p}$, the per-trajectory loss is
\begin{equation}
    \mathcal{L}_k(\theta) := \sum_{s\in\mathcal{I}_k}\|x_s^+ - \Phi_s\theta\|_2^2.
    \label{eq:traj_loss}
\end{equation}
The model is estimated using Tikhonov-regularized least squares:
\begin{equation}
    \hat{\theta} = \argmin_{\theta\in\R^p} L_{\lambda}(\theta,\mathcal{D}),
    \quad
    L_{\lambda}(\theta,\mathcal{D}) := \sum_{k=1}^N \mathcal{L}_k(\theta) + \lambda\|\theta\|_2^2,
    \label{eq:reg_obj}
\end{equation}
with $\lambda > 0$, ensuring that $\hat{\theta}$ is unique.

\textbf{Control design stage.} Given the identified model $(A_{\hat{\theta}},B_{\hat{\theta}})$, the infinite-horizon discrete-time LQR gain is obtained from the stabilizing solution of the DARE:
\begin{equation}
    P = Q + A^{\top}\!PA - A^{\top}\!PB(R+B^{\top}\!PB)^{-1}B^{\top}\!PA,
    \label{eq:dare}
\end{equation}
where we write $A = A_{\theta}$, $B = B_{\theta}$ for brevity, $Q \succeq 0$, and $R \succ 0$. The nominal LQR design metric is
\begin{equation}
    J(\theta;\Sigma_0) := \tr(P(\theta)\Sigma_0), \quad \Sigma_0 \succeq 0.
    \label{eq:J_def}
\end{equation}

\textbf{Goal.} Given a trajectory $\tau_k$, we seek to approximate the effect of removing it from $\mathcal{D}$ on (i)~the parameter vector, (ii)~a held-out prediction loss, and (iii)~the nominal LQR metric---without retraining.

\section{LOTO Analysis for Identification}

\subsection{Exact Parameter Shift}

Let $\mathcal{D}_{\setminus k} := \mathcal{D}\setminus\{\tau_k\}$ and $\hat{\theta}_{\setminus k} := \argmin_{\theta} L_{\lambda}(\theta,\mathcal{D}_{\setminus k})$. Define the LOTO parameter shift $\Delta\hat{\theta}_k := \hat{\theta}_{\setminus k} - \hat{\theta}$ and the quantities
\begin{equation}
    g_k := \nabla \mathcal{L}_k(\hat{\theta}), \quad
    H := \nabla^2 L_{\lambda}(\hat{\theta},\mathcal{D}), \quad
    H_k := \nabla^2 \mathcal{L}_k(\hat{\theta}).
    \label{eq:gH_defs}
\end{equation}
Since $L_{\lambda}$ is quadratic in $\theta$, the Hessians are constant:
\begin{equation}
    H = 2\!\sum_{i=1}^{N}\sum_{s\in\mathcal{I}_i}\!\Phi_s^{\top}\Phi_s + 2\lambda I, \quad
    H_k = 2\!\sum_{s\in\mathcal{I}_k}\!\Phi_s^{\top}\Phi_s.
    \label{eq:H_Hk}
\end{equation}

\begin{prop}[Exact LOTO shift]\label{prop:exact_loto}
If $H - H_k \succ 0$, then
\begin{equation}
    \Delta\hat{\theta}_k = (H-H_k)^{-1} g_k.
    \label{eq:exact_loto}
\end{equation}
\end{prop}

\begin{pf}
The optimality conditions $\nabla L_{\lambda}(\hat{\theta},\mathcal{D}) = 0$ and $\nabla L_{\lambda}(\hat{\theta}_{\setminus k},\mathcal{D}_{\setminus k}) = 0$, together with the identity $L_{\lambda}(\theta,\mathcal{D}_{\setminus k}) = L_{\lambda}(\theta,\mathcal{D}) - \mathcal{L}_k(\theta)$ and the fact that all gradients are affine in $\theta$ (since the objectives are quadratic), yield
\[
H\Delta\hat{\theta}_k = g_k + H_k\Delta\hat{\theta}_k,
\]
from which \eqref{eq:exact_loto} follows. The condition $H - H_k \succ 0$ holds whenever $\lambda > 0$ and $\sum_{i \neq k} T_i \geq p/n_x$, which is satisfied in all practical regimes.
\end{pf}

\begin{remark}[Sufficient condition for invertibility]\label{rem:invert}
Since $H - H_k = 2\sum_{i \neq k}\sum_{s \in \mathcal{I}_i} \Phi_s^\top \Phi_s + 2\lambda I$ and $\lambda > 0$, the matrix $H - H_k$ is always positive definite. The exact shift \eqref{eq:exact_loto} therefore holds for any $\tau_k$ without further assumptions.
\end{remark}

\subsection{Reusable First-Order Approximation}

\begin{prop}[Influence approximation and error bound]\label{prop:approx}
Define $\delta_k := \|H^{-1/2}H_kH^{-1/2}\|_2$. If $\delta_k < 1$, then
\begin{equation}
    \Delta\hat{\theta}_k = H^{-1}g_k + H^{-1}H_kH^{-1}g_k + r_k,
    \label{eq:neumann_expansion}
\end{equation}
where $\|r_k\| \leq \frac{\delta_k^2}{1-\delta_k}\|H^{-1}g_k\|$. Hence the first-order approximation
\begin{equation}
    \Delta\hat{\theta}_k \approx H^{-1}g_k
    \label{eq:first_order}
\end{equation}
has relative error bounded by $\delta_k/(1-\delta_k)$.
\end{prop}

\begin{pf}
Write $(H-H_k)^{-1} = (I - H^{-1}H_k)^{-1}H^{-1}$. Since $\delta_k = \|H^{-1/2}H_kH^{-1/2}\|_2 < 1$, the Neumann series $\sum_{j=0}^{\infty}(H^{-1}H_k)^j$ converges. The zeroth-order term gives \eqref{eq:first_order}; the first-order correction is $H^{-1}H_kH^{-1}g_k$; the remainder $r_k$ satisfies $\|r_k\| \leq \sum_{j=2}^{\infty}\delta_k^j \|H^{-1}g_k\| = \frac{\delta_k^2}{1-\delta_k}\|H^{-1}g_k\|$.
\end{pf}

The quantity $\delta_k$ captures the curvature contribution of trajectory $\tau_k$ relative to the full dataset. When $N$ is large and trajectories are comparably informative, $\delta_k \approx 1/N$.

\subsection{Identification-Level Influence Score}

Let $L_{\mathrm{pred}}(\theta)$ denote a held-out prediction loss evaluated at parameters $\theta$. When $L_{\mathrm{pred}}$ is the squared prediction error on test data, it too is quadratic, and the LOTO effect decomposes exactly as
\begin{equation}
    L_{\mathrm{pred}}(\hat{\theta}_{\setminus k}) - L_{\mathrm{pred}}(\hat{\theta})
    = \nabla L_{\mathrm{pred}}(\hat{\theta})^{\top}\Delta\hat{\theta}_k + \tfrac{1}{2}\Delta\hat{\theta}_k^{\top}H_{\mathrm{pred}}\Delta\hat{\theta}_k.
    \label{eq:pred_exact}
\end{equation}
The first-order influence score for identification is then
\begin{equation}
    \mathrm{IF1}_k := g_k^{\top}H^{-1}\nabla L_{\mathrm{pred}}(\hat{\theta}).
    \label{eq:IF1}
\end{equation}

\section{Influence Propagation Through the DARE}

\subsection{Setup and Notation}

Denote the nominal quantities at $\hat{\theta}$ by $A_0 := A_{\hat{\theta}}$, $B_0 := B_{\hat{\theta}}$, $P_0 := P(\hat{\theta})$, and define
\begin{equation}
\begin{aligned}
M_0 &:= R + B_0^{\top} P_0 B_0, \\
K_0 &:= M_0^{-1} B_0^{\top} P_0 A_0, \\
A_{\mathrm{cl}} &:= A_0 - B_0 K_0 .
\end{aligned}
\label{eq:cl_defs}
\end{equation}

\begin{assumption}\label{ass:riccati}
At $\theta = \hat{\theta}$: (i) $(A_0,B_0)$ is stabilizable; (ii) $(A_0,Q^{1/2})$ is detectable; (iii) $A_{\mathrm{cl}}$ is Schur stable (i.e., $\rho(A_{\mathrm{cl}}) < 1$).
\end{assumption}

Under Assumption~\ref{ass:riccati}, the DARE \eqref{eq:dare} admits a unique stabilizing solution $P_0 \succeq 0$ \cite{lancaster1995algebraic}.

\subsection{Fr\'{e}chet Derivative of the DARE Residual}

Define the DARE residual map $\mathcal{R}: \mathbb{S}^{n_x} \times \R^p \to \mathbb{S}^{n_x}$:
\begin{equation}
    \mathcal{R}(P,\theta) := P - Q - A^{\top}\!PA + A^{\top}\!PB(R+B^{\top}\!PB)^{-1}B^{\top}\!PA.
    \label{eq:dare_residual}
\end{equation}

\begin{prop}[Fr\'{e}chet derivative with respect to $P$]\label{prop:frechet}
The Fr\'{e}chet derivative of $\mathcal{R}$ with respect to $P$ at $(P_0, \hat{\theta})$, acting on a symmetric perturbation $\Delta P \in \mathbb{S}^{n_x}$, is
\begin{equation}
    D_P\mathcal{R}[P_0,\hat{\theta}](\Delta P) = \Delta P - A_{\mathrm{cl}}^{\top}\,\Delta P\, A_{\mathrm{cl}}.
    \label{eq:DPR}
\end{equation}
\end{prop}

\begin{pf}
We compute $\frac{d}{d\epsilon}\mathcal{R}(P_0 + \epsilon\Delta P, \hat{\theta})\big|_{\epsilon=0}$. Define $F(P) := A_0^{\top}PB_0(R + B_0^{\top}PB_0)^{-1}B_0^{\top}PA_0$. The derivative of the inverse term is $\frac{d}{d\epsilon}(R + B_0^{\top}(P_0+\epsilon\Delta P)B_0)^{-1}\big|_{\epsilon=0} = -M_0^{-1}B_0^{\top}\Delta P\, B_0 M_0^{-1}$. Applying the product rule to $F$ and substituting $K_0 = M_0^{-1}B_0^{\top}P_0A_0$:
\begin{align}
    \frac{dF}{d\epsilon}\Big|_{\epsilon=0}
    &= A_0^{\top}\Delta P\, B_0 K_0
    + K_0^{\top} B_0^{\top}\Delta P\, A_0 \nonumber\\
    &\quad - K_0^{\top}B_0^{\top}\Delta P\, B_0 K_0.
    \label{eq:dF_simplified}
\end{align}
The key algebraic identity is
\begin{align}
    &A_0^{\top}\Delta P\, B_0 K_0 + K_0^{\top} B_0^{\top}\Delta P\, A_0 - K_0^{\top}B_0^{\top}\Delta P\, B_0 K_0 \nonumber\\
    &= A_0^{\top}\Delta P\, A_0 - (A_0 - B_0K_0)^{\top}\Delta P\,(A_0 - B_0K_0) \nonumber\\
    &= A_0^{\top}\Delta P\, A_0 - A_{\mathrm{cl}}^{\top}\Delta P\, A_{\mathrm{cl}},
    \label{eq:key_identity}
\end{align}
which can be verified by expanding $(A_0 - B_0K_0)^{\top}\Delta P\,(A_0 - B_0K_0)$. Therefore,
\[
D_P\mathcal{R}(\Delta P) = \Delta P - A_0^{\top}\Delta P\, A_0 + \frac{dF}{d\epsilon}\Big|_{\epsilon=0} = \Delta P - A_{\mathrm{cl}}^{\top}\Delta P\, A_{\mathrm{cl}}.
\]
\end{pf}

\begin{remark}
Since $A_{\mathrm{cl}}$ is Schur stable under Assumption~\ref{ass:riccati}, the linear operator $\mathcal{T}(\Delta P) := \Delta P - A_{\mathrm{cl}}^{\top}\Delta P\, A_{\mathrm{cl}}$ is invertible on $\mathbb{S}^{n_x}$: its eigenvalues take the form $1 - \bar{\mu}_i\mu_j$ where $\mu_i$ are eigenvalues of $A_{\mathrm{cl}}$, all satisfying $|\mu_i| < 1$. This is the discrete Lyapunov operator.
\end{remark}

\subsection{Riccati Sensitivity via Implicit Differentiation}

Since $\theta = \vecop(A_{\theta}, B_{\theta})$, each coordinate $\theta_m$ corresponds to an entry of either $A_{\theta}$ or $B_{\theta}$. Let $\mathcal{A}_m := \partial A_{\theta}/\partial\theta_m|_{\hat{\theta}}$ and $\mathcal{B}_m := \partial B_{\theta}/\partial\theta_m|_{\hat{\theta}}$; these are standard basis matrices (a single entry equal to one).

\begin{prop}[Riccati sensitivity]\label{prop:riccati_sens}
Under Assumption~\ref{ass:riccati}, $P(\theta)$ is locally differentiable at $\hat{\theta}$, and the sensitivity matrix $S_m := \partial P/\partial\theta_m(\hat{\theta})$ is the unique symmetric solution of the discrete Lyapunov equation
\begin{equation}
    S_m - A_{\mathrm{cl}}^{\top}S_m A_{\mathrm{cl}} = -\frac{\partial \mathcal{R}}{\partial\theta_m}(P_0,\hat{\theta}),
    \label{eq:lyap_sens}
\end{equation}
where the right-hand side is
\begin{align}
    \frac{\partial \mathcal{R}}{\partial\theta_m}
    &= -(\mathcal{A}_m - \mathcal{B}_mK_0)^{\top}P_0(A_0 - B_0K_0) \nonumber\\
    &\quad - (A_0 - B_0K_0)^{\top}P_0(\mathcal{A}_m - \mathcal{B}_mK_0) \nonumber\\
    &\quad - K_0^{\top}\mathcal{B}_m^{\top}P_0\mathcal{B}_mK_0  \nonumber\\
    &\quad + K_0^{\top}\mathcal{B}_m^{\top}P_0(A_0 - B_0K_0) \nonumber\\
    &\quad + (A_0 - B_0K_0)^{\top}P_0\mathcal{B}_mK_0.
    \label{eq:dR_dtheta_compact}
\end{align}
\end{prop}

\begin{pf}
Differentiating $\mathcal{R}(P(\theta),\theta) = 0$ with respect to $\theta_m$ and applying the chain rule gives
\begin{equation}
    D_P\mathcal{R}[P_0,\hat{\theta}](S_m) + \frac{\partial\mathcal{R}}{\partial\theta_m}(P_0,\hat{\theta}) = 0.
    \label{eq:ift}
\end{equation}
By Proposition~\ref{prop:frechet}, $D_P\mathcal{R}(S_m) = S_m - A_{\mathrm{cl}}^{\top}S_mA_{\mathrm{cl}}$. Rearranging yields \eqref{eq:lyap_sens}. Local differentiability of $P(\theta)$ follows from the implicit function theorem, since $D_P\mathcal{R}$ is invertible under Assumption~\ref{ass:riccati}.

For \eqref{eq:dR_dtheta_compact}, we differentiate $\mathcal{R}(P_0,\theta)$ with respect to $\theta_m$ at $\hat{\theta}$, holding $P = P_0$ fixed. Rewriting the DARE in closed-loop form $P_0 = Q + A_{\mathrm{cl}}^{\top}P_0A_{\mathrm{cl}} + K_0^{\top}M_0K_0$ and differentiating with respect to $\theta_m$---noting that $K_0$ depends on $P_0$, which is held fixed---gives the compact expression
\begin{equation}
    \frac{\partial\mathcal{R}}{\partial\theta_m} = -\mathcal{A}_{\mathrm{cl},m}^{\top}P_0A_{\mathrm{cl}} - A_{\mathrm{cl}}^{\top}P_0\mathcal{A}_{\mathrm{cl},m},
    \label{eq:dR_compact}
\end{equation}
where $\mathcal{A}_{\mathrm{cl},m} := \mathcal{A}_m - \mathcal{B}_mK_0$ is the sensitivity of the closed-loop matrix to $\theta_m$. This follows because differentiating $A_{\mathrm{cl}}^{\top}P_0A_{\mathrm{cl}}$ with respect to $\theta_m$ produces $\mathcal{A}_{\mathrm{cl},m}^{\top}P_0A_{\mathrm{cl}} + A_{\mathrm{cl}}^{\top}P_0\mathcal{A}_{\mathrm{cl},m}$, while the $K_0^{\top}M_0K_0$ term generates additional contributions that cancel upon substituting the definition of $K_0$. Expanding \eqref{eq:dR_compact} recovers \eqref{eq:dR_dtheta_compact}.
\end{pf}

\subsection{End-to-End Influence Score via Adjoint Sensitivity}

The naive approach to computing $\nabla J$ solves $p$ forward Lyapunov equations \eqref{eq:lyap_sens}—one per parameter coordinate—obtains $S_m$, and forms $\partial J/\partial\theta_m = \tr(S_m\Sigma_0)$. Since $p = n_x(n_x+n_u)$, this rapidly becomes the computational bottleneck. An adjoint reformulation sidesteps it entirely with a single Lyapunov solve.

Define the adjoint operator $\mathcal{T}^{*}: \mathbb{S}^{n_x} \to \mathbb{S}^{n_x}$ by $\mathcal{T}^{*}(Y) := Y - A_{\mathrm{cl}}\,Y\,A_{\mathrm{cl}}^{\top}$; the adjoint of $\mathcal{T}$ under the Frobenius inner product $\langle X, Y \rangle := \tr(X^{\top}Y)$.

\begin{prop}[Adjoint gradient computation]\label{prop:adjoint}
Under Assumption~\ref{ass:riccati}, let $\Lambda \in \mathbb{S}^{n_x}$ be the unique solution of the adjoint Lyapunov equation
\begin{equation}
    \Lambda - A_{\mathrm{cl}}\,\Lambda\, A_{\mathrm{cl}}^{\top} = \Sigma_0.
    \label{eq:adjoint_lyap}
\end{equation}
Then the gradient of $J(\theta;\Sigma_0) = \tr(P(\theta)\Sigma_0)$ is
\begin{equation}
    \frac{\partial J}{\partial\theta_m} = -\tr\!\left(\Lambda\,\frac{\partial\mathcal{R}}{\partial\theta_m}(P_0,\hat{\theta})\right),
    \label{eq:J_grad_adjoint}
\end{equation}
where $\partial\mathcal{R}/\partial\theta_m$ is given by \eqref{eq:dR_compact}.
\end{prop}

\begin{pf}
From Proposition~\ref{prop:riccati_sens}, $\mathcal{T}(S_m) = -\partial\mathcal{R}/\partial\theta_m$. Therefore,
\begin{align}
    \frac{\partial J}{\partial\theta_m}
    &= \tr(S_m\,\Sigma_0)
    = \langle S_m,\,\Sigma_0 \rangle
    = \langle S_m,\,\mathcal{T}^{*}(\Lambda) \rangle \nonumber\\
    &= \langle \mathcal{T}(S_m),\,\Lambda \rangle
    = -\tr\!\left(\Lambda\,\frac{\partial\mathcal{R}}{\partial\theta_m}\right),
    \label{eq:adjoint_derivation}
\end{align}
where the third equality uses $\Sigma_0 = \mathcal{T}^{*}(\Lambda)$ from \eqref{eq:adjoint_lyap}, and the fourth invokes the adjointness relation $\langle \mathcal{T}(X), Y \rangle = \langle X, \mathcal{T}^{*}(Y) \rangle$.
\end{pf}

\begin{remark}[Interpretation]
The matrix $\Lambda$ is the reachability Gramian of the closed-loop system with $\Sigma_0$ as the driving term. It plays a role analogous to the costate in optimal control, converting DARE residual perturbations into scalar cost perturbations and eliminating the need to compute the full sensitivity matrices $S_m$.
\end{remark}

Substituting \eqref{eq:dR_compact} into \eqref{eq:J_grad_adjoint}, each gradient component reduces to
\begin{equation}
    \frac{\partial J}{\partial\theta_m} = \tr\!\bigl(\Lambda\bigl[\mathcal{A}_{\mathrm{cl},m}^{\top}P_0A_{\mathrm{cl}} + A_{\mathrm{cl}}^{\top}P_0\mathcal{A}_{\mathrm{cl},m}\bigr]\bigr),
    \label{eq:J_grad_expanded}
\end{equation}
a trace of matrix products involving $\Lambda$, $P_0$, and $A_{\mathrm{cl}}$, evaluated at each standard-basis perturbation $\mathcal{A}_{\mathrm{cl},m}$.

Assembling the full gradient $\nabla J(\hat{\theta}) \in \R^p$, the control-level influence score for trajectory $\tau_k$ is
\begin{equation}
    \mathrm{IF2}_k := g_k^{\top}H^{-1}\nabla J(\hat{\theta}),
    \label{eq:IF2}
\end{equation}
which approximates $J(\hat{\theta}_{\setminus k};\Sigma_0) - J(\hat{\theta};\Sigma_0)$ to first order.

\begin{cor}[Computational cost]\label{cor:cost}
Under the adjoint formulation, computing all $N$ scores $\{\mathrm{IF2}_k\}_{k=1}^N$ requires one Cholesky factorization of $H$ ($O(p^3)$), one Lyapunov solve for $\Lambda$ ($O(n_x^3)$), $p$ trace products to assemble $\nabla J$ ($O(p\cdot n_x^2)$), and $N$ dot products ($O(Np)$). The total cost is $O(p^3 + n_x^3 + p\cdot n_x^2 + Np)$, compared with $O(p^3 + p\cdot n_x^3 + Np)$ for forward sensitivity and $O(Np^3)$ for full LOTO retraining.
\end{cor}

\begin{algorithm}[t]
\caption{End-to-End Trajectory Influence Scores}
\label{alg:influence}
\begin{algorithmic}[1]
\REQUIRE Dataset $\mathcal{D}$, regularization $\lambda$, cost matrices $Q,R,\Sigma_0$
\STATE Solve \eqref{eq:reg_obj} for $\hat{\theta}$; extract $A_0, B_0$
\STATE Compute $H$ via \eqref{eq:H_Hk}; factorize $H = LL^{\top}$
\STATE Solve DARE \eqref{eq:dare} for $P_0$; compute $K_0, A_{\mathrm{cl}}$
\STATE Solve adjoint Lyapunov equation \eqref{eq:adjoint_lyap} for $\Lambda$
\FOR{$m = 1,\ldots,p$}
    \STATE Compute $\mathcal{A}_{\mathrm{cl},m} = \mathcal{A}_m - \mathcal{B}_mK_0$
    \STATE $[\nabla J]_m \leftarrow \tr\!\bigl(\Lambda\bigl[\mathcal{A}_{\mathrm{cl},m}^{\top}P_0A_{\mathrm{cl}} + A_{\mathrm{cl}}^{\top}P_0\mathcal{A}_{\mathrm{cl},m}\bigr]\bigr)$
\ENDFOR
\STATE $v \leftarrow H^{-1}\nabla J(\hat{\theta})$ \hfill (via Cholesky back-solve)
\FOR{$k = 1,\ldots,N$}
    \STATE $g_k \leftarrow -2\sum_{s\in\mathcal{I}_k}\Phi_s^{\top}(x_s^+ - \Phi_s\hat{\theta})$
    \STATE $\mathrm{IF1}_k \leftarrow g_k^{\top}H^{-1}\nabla L_{\mathrm{pred}}(\hat{\theta})$
    \STATE $\mathrm{IF2}_k \leftarrow g_k^{\top}v$
\ENDFOR
\RETURN $\{\mathrm{IF1}_k, \mathrm{IF2}_k\}_{k=1}^N$
\end{algorithmic}
\end{algorithm}

\section{Approximation Error Analysis}

The end-to-end score $\mathrm{IF2}_k$ combines two layers of approximation—one from the LOTO parameter estimate and one from linearizing the Riccati map—and each has identifiable failure modes.

LOTO approximation. By Proposition~\ref{prop:approx}, replacing the exact shift with $H^{-1}g_k$ incurs a relative error of order $\delta_k = \|H^{-1/2}H_kH^{-1/2}\|_2$. This remains small whenever each trajectory contributes only a modest share of the total curvature—a condition caused by large $N$, moderate trajectory lengths, and sufficient regularization.

Riccati linearization. The first-order expansion $J(\hat{\theta}+\Delta) \approx J(\hat{\theta}) + \nabla J^{\top}\Delta$ leaves a remainder $\frac{1}{2}\Delta^{\top}\nabla^2 J(\tilde{\theta})\Delta$. The Hessian $\nabla^2 J$, which involves second derivatives of the DARE solution, can grow large in three regimes: when $A_{\mathrm{cl}}$ is close to the unit circle ($\rho(A_{\mathrm{cl}}) \to 1$), amplifying the Lyapunov operator's condition number; when the parameter perturbation $\|\Delta\hat{\theta}_k\|$ is large, as occurs with a dominant trajectory or small $N$.

Nominal vs.\ plant-level metrics. Both $\mathrm{IF1}_k$ and $\mathrm{IF2}_k$ target \emph{nominal} quantities: the surrogate prediction loss and the DARE-based design cost. When the true plant is nonlinear, these nominal metrics can diverge from the actual closed-loop cost. That gap is a modeling issue rather than an influence-approximation issue, and it widens with model mismatch. The experiments below measure both nominal and plant-level metrics to separate these two sources of error.

\section{Empirical Evaluation}

We evaluate the proposed scores on four systems of increasing complexity.

\textbf{Systems.} S1 is a small, stable LTI system ($n_x=2$, $n_u=1$) based on a mass–spring–damper, included primarily as a sanity check. S2 is a medium-sized, stable LTI system ($n_x=4$, $n_u=2$) inspired by a planar vehicle with lateral and longitudinal dynamics. S3 is a higher-dimensional, stable LTI family ($n_x\in\{8,10\}$, $n_u\in\{3,4\}$) drawn from coupled multi-body systems, used to probe scaling behavior. S4 is a nonlinear two-link manipulator ($n_x=4$, $n_u=2$) for which we fit a local linear surrogate; influence is computed on the surrogate, but the closed-loop cost is evaluated on the true nonlinear plant.

\textbf{Setup.} We use $N=30$ trajectories for S1, $N=50$ for S2 and S4, and $N=80$ for S3, with a trajectory length of $T=25$ for S1 and $T=30$ otherwise. The remaining parameters are $\sigma_w=0.03$, $\lambda=10^{-5}$, $Q=I$, $R=0.1I$, and $\Sigma_0=I$; results are averaged over multiple seeds. Inverse-Hessian-vector products are computed via conjugate gradient (CG) following \cite{koh2017understanding} rather than by forming $H^{-1}$ explicitly. On S2, replacing the exact structured solve with CG leaves IF1 and IF2 accuracy unchanged to the displayed precision—the choice primarily affects runtime, not statistical quality. The ablation sweeps vary $N$, $T$, $\sigma_w$, $\lambda$, closed-loop conditioning, and model mismatch.

\textbf{Baselines.} We compare against: (i) exact LOTO via \eqref{eq:exact_loto}; (ii) a gradient-only score, using \(g_k^{\top}\nabla L_{\mathrm{pred}}(\hat{\theta})\) for prediction targets and \(g_k^{\top}\nabla J(\hat{\theta})\) for control targets; and (iii) the residual norm \(\mathcal{L}_k(\hat{\theta})\).

\textbf{Metrics.} We report Pearson and Spearman correlation, MAE, and top-\(k\) overlap with \(k=5\) between predicted and true LOTO effects.

\subsection{LOTO Prediction Accuracy}

\begin{figure*}[t]
    \centering
    \includegraphics[width=\textwidth]{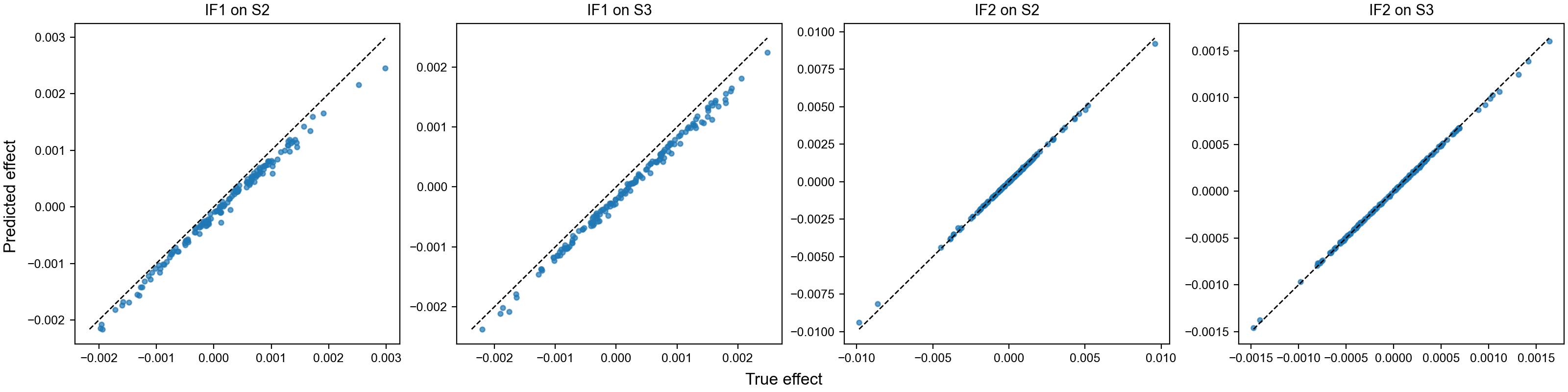}
\caption{Predicted versus true LOTO effects on the linear benchmarks (S1--S3).}
\label{fig:main_scatter}
\end{figure*}

Table~\ref{tab:main_results} summarizes the main findings, and Fig.~\ref{fig:main_scatter} plots predicted versus true LOTO effects. On the linear systems S1 and S2, influence scores perform well: IF1 achieves Pearson correlations of $0.995$ (S1) and $0.997$ (S2) for held-out prediction loss, while IF2 reaches $0.999$ (S1) and effectively $1.000$ (S2) for the nominal LQR metric. The exact quadratic LOTO formula matches explicit retraining to machine precision. Gradient-only baselines are noticeably weaker, and residual-norm ranking carries almost no information about true LOTO effects on these benchmarks. At the larger scale of S3, accuracy degrades only marginally—IF1 at $0.993$ and IF2 at $0.998$—confirming that the approximation holds as the parameter count grows.

The nonlinear S4 results are less encouraging. With the controller evaluated on the true plant, IF2 is only moderately predictive (Pearson $0.620$, Spearman $0.512$, top-$k$ overlap $0.250$).

The problem is not the influence approximation but the gap between the surrogate and the real plant. The two-link manipulator has configuration-dependent inertia and Coriolis terms that a single linear model cannot capture, so the surrogate treats these nonlinearities as noise. When a trajectory wanders far from the linearization point, the resulting parameter shift may help the surrogate locally while hurting closed-loop performance on the actual plant— which explains the low top-$k$ overlap. A simple check supports this: evaluating IF2 against the \emph{nominal} surrogate cost instead of the plant-level cost brings the correlation back up, consistent with the S2 mismatch ablation.

Still, a Pearson correlation of $0.620$ is not without value. The scores carry enough signal to serve as a coarse screen—flagging trajectories worth a closer look, even if the fine-grained ranking is unreliable. One way to improve them would be to weight each trajectory's influence by a measure of local linearization quality, discounting scores where the surrogate residuals show systematic patterns rather than i.i.d. noise. We leave this for future work.

\subsection{Computational Scaling}

Table~\ref{tab:main_results} also reports runtimes. On S3, the largest benchmark, IF1 is $13\times$ faster than full LOTO retraining and IF2 is $7.2\times$ faster, while exact quadratic LOTO yields a more modest $3.4\times$ speedup. IF1 and IF2 outperform exact LOTO because of amortization: they factorize the Hessian once and reduce each per-trajectory computation to a dot product, whereas exact LOTO requires solving $(H - H_k)^{-1}g_k$ separately for every trajectory. This pattern persists across the smaller systems, with even larger gains on S1 and S2 where the Hessian factorization cost is negligible. The runtime gap between IF1 and IF2 reflects the forward sensitivity approach used in these experiments, which solves one Lyapunov equation per parameter direction. The adjoint formulation of Proposition~\ref{prop:adjoint}, implemented in Algorithm~\ref{alg:influence}, replaces all $p$ solves with a single one, reducing the DARE sensitivity cost from $O(p\cdot n_x^3)$ to $O(n_x^3 + p\cdot n_x^2)$. For S3 ($n_x=10$, $n_u=4$, $p=140$), this eliminates the dominant component of IF2's cost and brings it in line with IF1.

\subsection{Approximation Quality Under Predicted Failure Regimes}

The ablation results align with the theory. On S1, IF1 holds up well across sweeps of dataset size, noise level, and regularization: Pearson correlations stay between $0.983$ and $0.997$ as $N$ varies, and between $0.992$ and $0.997$ across $\lambda\in\{10^{-7},10^{-5},10^{-3}\}$.

IF2 on S2 is even more stable — the Pearson correlation sits between $0.9996$ and $0.9999$ as the target spectral radius goes from $0.70$ to $0.95$, meaning the Riccati linearization barely matters in this range. Where things break down is model mismatch. The correlation between IF2 and true plant-level performance drops from $0.011$ at zero mismatch to roughly $-0.209$ and $-0.186$ at mismatch strengths $0.02$ and $0.05$. The same pattern shows up in S4: the weak link is not the inverse-Hessian approximation or the DARE sensitivity, but the gap between what the surrogate says and what the real plant does.

\begin{table*}[t]
\caption{LOTO prediction accuracy and computational cost. Correlations are against true LOTO retraining; speedup is relative to LOTO retraining within each system block.}
\label{tab:main_results}
\centering
\footnotesize
\setlength{\tabcolsep}{4pt}
\renewcommand{\arraystretch}{0.93}
\begin{tabular}{@{}lll cccc cc@{}}
\toprule
Sys.\ & Target & Method & Pears.\ & Spear.\ & MAE & Top-$k$ & Time (s) & Speedup \\
\midrule
S1 & Pred.\ loss  & Residual    & $-$0.059 & $-$0.037 & 0.043            & 0.000 & 1.2e-4 & $175\times$ \\
   & Pred.\ loss  & Grad.-only  & 0.875    & 0.873    & 0.833            & 0.667 & 1.8e-4 & $117\times$ \\
   & Pred.\ loss  & IF1         & 0.995    & 0.997    & $1.1\text{e-}4$  & 0.667 & 3.5e-4 & $60\times$ \\
   & LQR metric   & IF2         & 0.999    & 1.000    & $1.4\text{e-}4$  & 0.667 & 5.8e-4 & $36\times$ \\
   & Pred.\ loss  & Exact LOTO  & 1.000    & 1.000    & $6.1\text{e-}16$ & 1.000 & 2.1e-3 & $10\times$ \\
   & --           & Retraining  & \multicolumn{4}{c}{\emph{ground truth}} & 2.1e-2 & $1\times$ \\
\midrule
S2 & LQR metric   & Residual    & 0.098 & 0.103 & 0.107            & 0.000 & 2.8e-4 & $171\times$ \\
   & LQR metric   & Grad.-only  & 0.647 & 0.737 & 2.640            & 0.250 & 5.2e-4 & $92\times$ \\
   & Pred.\ loss  & IF1         & 0.997 & 0.996 & $1.6\text{e-}4$  & 0.667 & 1.8e-3 & $27\times$ \\
   & LQR metric   & IF2         & 1.000 & 1.000 & $3.7\text{e-}5$  & 1.000 & 3.5e-3 & $14\times$ \\
   & --           & Retraining  & \multicolumn{4}{c}{\emph{ground truth}} & 4.8e-2 & $1\times$ \\
\midrule
S3 & Pred.\ loss  & IF1         & 0.993 & 0.988 & $3.1\text{e-}4$  & 0.800 & 8.2e-3 & $13\times$ \\
   & LQR metric   & IF2         & 0.998 & 0.996 & $7.5\text{e-}5$  & 0.800 & 1.5e-2 & $7.2\times$ \\
   & Pred.\ loss  & Exact LOTO  & 1.000 & 1.000 & $8.3\text{e-}16$ & 1.000 & 3.1e-2 & $3.4\times$ \\
   & --           & Retraining  & \multicolumn{4}{c}{\emph{ground truth}} & 1.0e-1 & $1\times$ \\
\midrule
S4 & Plant cost   & Grad.-only  & 0.521 & 0.413 & 46.28            & 0.111 & 5.5e-4 & $84\times$ \\
   & Plant cost   & IF2         & 0.620 & 0.512 & 0.142            & 0.250 & 3.8e-3 & $12\times$ \\
   & --           & Retraining  & \multicolumn{4}{c}{\emph{ground truth}} & 4.6e-2 & $1\times$ \\
\bottomrule
\end{tabular}
\end{table*}

\section{Conclusion and Future Work}

This paper extends influence functions to trajectory-level data attribution in identification-to-LQR pipelines, chaining closed-form least-squares influence with adjoint Lyapunov-based DARE sensitivity. On linear systems of dimension $2$--$10$, the scores closely match true LOTO retraining (Pearson $\geq 0.99$) at $7$--$60\times$ the speed, making them practical for dataset auditing and controller debugging. On a nonlinear two-link manipulator with a local linear surrogate, the scores are still moderately informative—the main source of error is the surrogate-to-plant gap, not the influence approximation itself.

\textbf{Future work:} There are several natural next steps. Influence scores could help diagnose distributional shifts between training and deployment data or quantify how individual trajectories affect safety-critical metrics like constraint satisfaction in MPC—though the latter would require differentiating through a parametric QP rather than the DARE. On the data-collection side, the scores could steer an agent toward regions where new trajectories would most improve controller performance. Further out, extending these ideas to coupled multi-agent settings, where one subsystem's data affects its neighbors through dynamic coupling.

\begin{ack}
Claude was used to assist with the language editing of this manuscript.
\end{ack}

\section*{DECLARATION OF GENERATIVE AI AND AI-ASSISTED TECHNOLOGIES IN THE WRITING PROCESS}
During the preparation of this work the author(s) used Claude in order to assist with the language editing of this manuscript. 


\bibliography{ifacconf}

\end{document}